# Spectral profile of ro-vibrational transitions of HCl broadened by He, Ar and SF$_6$: testing the *β*-correction to the Hartmann-Tran profile and the speed dependent (complex) hard collision model[1]

T. Le[1,#], J.-L. Domenech[2], N.H. Ngo[1], H. Tran[3, #]

[1] Faculty of Physics, Hanoi National University of Education, 136 Xuan Thuy Street, Cau Giay District, Hanoi, Vietnam

[2] Instituto de Estructura de la Materia, Consejo Superior de Investigaciones Cientificas (IEM-CSIC), Serrano 123, 28006 Madrid, Spain

[3] Laboratoire de Météorologie Dynamique/IPSL, CNRS, Sorbonne Universités, École normale supérieure, PSL Research University, École polytechnique, F-75005 Paris, France

[#] Corresponding author: tuonglc@hnue.edu.vn, ha.tran@lmd.jussieu.fr

**Abstract:**

The β-correction to the Hartmann-Tran (HT) profile, recently introduced to model the spectral shape of a molecular transition strongly affected by the Dicke narrowing effect (Konefal et al., JQSRT 242 (2020) 106784), the HT profile (Ngo et al., JQSRT 129 (2013) 89), and the speed dependent (complex) hard collision model (SDcHC, in which the velocity changing collision rate is characterized by a complex number) are tested using spectra of two rovibrational lines of HCl. The R(5) and R(9) lines of the fundamental band of HCl broadened by He, Ar and SF$_6$ have been recorded with a difference-frequency laser spectrometer for total pressures ranging from 12 to 930 mbar. These lines, measured in large pressure ranges with different collision-partners provide a meaningful test of the above-mentioned line-shape models. The results confirm that non-Voigt effects are significant for HCl broadened by Ar and SF$_6$ and mainly due to the large influence of the Dicke narrowing effect. For HCl-SF$_6$ and HCl-Ar, especially for the R(9) line, using the β-correction together with the HT profile (and with the speed-dependent hard collision model, SDHC) significantly improves the fit residuals while it has no effect on (or tends to deteriorate) the quality of the fit for HCl-He and for the R(5) line of HCl-SF$_6$, for which the influence of velocity changing collisions is smaller. Except for HCl-He, the SDcHC model leads to better quality of fit compared to the HT profile. The results also show that numerical correlations between refined line-shape parameters of the HT profile are important and can lead to ill-determined parameters while they are more properly determined with the SDcHC model.

**Keywords:** HCl, line shape, Hartmann-Tran profile, *β*-correction, complex Dicke narrowing parameter, difference frequency laser spectrometer

---

## 1. Introduction

The Hartmann-Tran (HT) profile [1] was recommended by IUPAC [2] as the reference line-shape model for high resolution spectroscopy, replacing the usual Voigt profile [3]. With its seven line-shape parameters, the HT profile takes into account several physical effects influencing the shape of an isolated line, i.e. (i) the Doppler broadening (characterized by the Doppler width $\Gamma_D$); (ii) the speed dependent collisional broadening (through the collisional line width $\Gamma_0$ and its quadratic speed dependent component $\Gamma_2$); (iii) the speed dependent line shift ($\Delta_0$ and $\Delta_2$); (iv) the collision-induced velocity changes ($\nu_{VC}$ – the velocity changing collision frequency parameter); and the correlation between velocity changing collisions and internal state changing collisions ($\eta$). This profile adopted the quadratic speed dependences of the line broadening and shifting [4,5] and the hard collision assumption [6] to model the collision-induced velocity changes effect. Contrary to the Voigt profile, the HT profile is capable to represent the measured line shapes of various molecular systems within a few 0.1% [1]. Importantly, unlike many other advanced line-shape models, the authors [1,7] showed that this profile can be expressed as a combination of Voigt functions, and therefore can be computed at low computer cost and easily used in radiative transfer codes. Since then, the HT model was successfully used to fit measured spectra of various molecular systems (e.g. [8-22]). The HITRAN spectroscopic database has been providing line-shape parameters associated with the HT profile for various molecules since 2016 [23,24], thanks to the new structure of the database. However, it was shown that for $H_2$ absorption spectra, for which the influence of collision-induced velocity changes effect is exceptionally pronounced [25-27] the HT model does not satisfactorily reproduce the measured line-shapes [27,28], leading to large residuals, up to 3% [1,28]. This behavior was attributed to the use of the hard-collision (HC) model in the HT profile to represent the velocity changes effect. Therefore, a correction to the HC approximation in the HT profile was proposed for pure $H_2$ [28] and then extended for other Dicke-narrowed molecular systems by Konefal et al. [29]. It was shown that the use of this so-called $\beta$-correction in the HT profile leads to very good agreement with measured spectra of pure $H_2$ for large ranges of pressure, in different bands and at different temperatures [28]. Note however, that a recent study of Odintsova and co-workers [30] showed that using this corrected profile for representing spectral shapes of pure $N_2O$ gas leads to the same results in comparison with the speed dependence hard collision model [1]. Another issue with the HT profile is the nontrivial way of evaluating the velocity changing collision rate and the correlation parameter for mixtures [1]. In addition, in Ref. [31], it was shown that the use of the HT profile leads to a singular behaviour of the temperature dependence of $\nu_{VC}$ and $\eta$ when the line shift crosses zero. The authors of Ref. [31] suggested to use the speed dependence (complex) hard collision (SDcHC) model to overcome these issues. This model can be easily obtained from the HT profile by setting the correlation parameter $\eta$ to zero and adding an imaginary part to the velocity changing collision frequency. Line-shape parameters of this model were obtained from ab initio quantum scattering calculations of generalized spectroscopic cross-sections for large temperature ranges, for $H_2$ and HD perturbed by He [32,33].

In this paper, the SDcHC, the HT profile and the β-correction are tested for the first time by comparison with spectra of the R(5) and R(9) lines in the fundamental band of HCl, broadened by He, Ar and $SF_6$, measured at room temperature and for broad pressure ranges. Because of its simple structure, large dipole moment and large rotational constant, HCl is of fundamental interest in the study of intermolecular interactions and a well suited system to test the different line-shape models for isolated transitions. The large Dicke narrowing effect on HCl spectra was reported for the first time by Wegdam and Sondag [34] who showed that, at some pressures, the observed width of the (1-0) P(15) line is significantly smaller than the Doppler width. Rao and Oka [35] observed similar phenomena for several lines of the (1-0) band of HCl diluted in Ar. In Refs. [36,37], asymmetries of the line shape were also observed.

Spectra of HCl lines, having strong Dicke narrowing effect, measured in large pressure ranges with different collision-partners thus provide a meaningful test of the $\beta$ -correction as well as the HT and SDcHC profiles. This paper is organized as follows: the measured spectra are described in the next section together with the analysis procedure. The obtained results are presented and discussed in Sec. 3 while the conclusions are drawn in Sec. 4.

## 2. Experimental data and spectra analysis
*2.1 Experimental setup and conditions*

Absorption spectra of the R(5) and R(9) in the fundamental band of HCl diluted in He, $SF_6$ and Ar were recorded with the difference-frequency laser spectrometer (DFLS) [38]. We used the same experimental setup as in Ref. [39], devoted to the measurements of Ar-broadened HCl spectra, where a detailed description can be found. The linewidth of the source is ~2 MHz, about 50 times smaller than the smallest apparent linewidth considered in this work. HCl gas was supplied by Praxair and the perturbers (He, Ar and $SF_6$) by Air Liquide. Since all considered gases had 99.9995% purity, they were used without further purification. Each mixture was prepared in a steel cylinder at least one day before experiments in order to get a homogeneous mixing. The mole fraction of HCl in each studied mixture was approximately 0.5%. Three cells with different absorption pathlengths were used in order to account for the about ten-time magnitude difference in intensity between the R(5) and R(9) transitions (see **Table 1**). The temperature of the cells wall was measured with a calibrated thermistor (±0.1 K) while the pressure was measured using two heated capacitance manometers with 1000 and 100 Torr full scale values, and accuracies ± 0.12% and ± 0.5% of the reading, respectively. The control unit resolution is 0.1 and 0.01 Torr, respectively.

For each collision-partner, absorption spectra of each transition were recorded at 10 different pressures (from 12.9 to 927.0 mbar) at room temperature (that ranged from 25 to $27^0$C but was constant within $0.1^0$C for each line, see **Table 1**). These pressures were chosen to cover from the Doppler regime to the collision-dominant range in order to involve contributions of different refined collisional effects on the spectral shape of HCl lines. Details of the experimental conditions for all measurements are presented in **Table 1**.

**Table 1**. Experimental conditions of the measured spectra. Uncertainties in pathlength and temperature are ±3σ estimations. The line positions and intensities were taken from the HITRAN database [24]. The spectra of HCl in Ar were selected from those of Ref. [39].

| Line | Frequency $\sigma_0$ (cm$^{-1}$) | Intensity (cm$^{-1}$/(molec.cm$^{-2}$)) | Perturber | Path length L (cm) | Temperature T (K) | Total pressure P (mbar) | $\Gamma/\Gamma_D$ | $P_{HCl}/P$ (%) |
|---|---|---|---|---|---|---|---|---|
| R(5) | 2998.046426 | 2.806 10$^{-19}$ | He | 4.0 ± 0.5 | 299 ± 1 | 684.4, 397.0, 251.1, 128.0, 79.6; | 0.07 - 3.87 | 0.497 |
| | | | | 10.0 ± 0.5 | 300 ± 1 | 66.8, 52.9, 40.0, 26.5, 13.1 | | |
| | | | SF$_6$ | 10.0 ± 0.5 | 300 ± 1 | 653.0, 400.5, 250.7, 129.2, 79.6, 66.9, 52.6, 40.0, 27.1, 13.0 | 0.17 - 8.72 | 0.530 |
| | | | Ar | 10.0 ± 0.5 | 298 ± 1 | 662.2, 399.8, 268.0, 134.5, 81.3, 66.7, 53.7, 40.4, 26.8, 13.5 | 0.09 - 4.37 | 0.444 |
| R(9) | 3059.316238 | 2.070 10$^{-20}$ | He | 24.2 ± 0.5 | 300 ± 1 | 795.8, 535.2, 399.8, 264.3, 200.2, 159.2, 119.3, 80.5, 40.1, 12.9 | 0.06 - 3.48 | 0.497 |
| | | | SF$_6$ | 24.2 ± 0.5 | 300 ± 1 | 651.8, 400.1, 250.1, 200.4, 150.5, 120.4, 90.5, 75.3, 59.8, 25.3 | 0.18 - 4.56 | 0.500 |
| | | | Ar | 24.2 ± 0.5 | 298 ± 1 | 927.0, 533.6, 400.0, 267.6, 199.6, 159.3, 119.6, 80.7, 53.4, 26.7 | 0.06 - 2.23 | 0.444 |

**Table 2**: Values of the β-correction function parameters for the considered molecular systems, determined using Eqs. (7a-d) of Ref. [29].

| System | α | $A_\alpha$ | $B_\alpha$ | $C_\alpha$ | $D_\alpha$ |
|---|---|---|---|---|---|
| HCl-He | 0.111182 | 0.204148 | 1.945589 | -0.047283 | 0.795852 |
| HCl-Ar | 1.110376 | 0.149460 | 1.833564 | 0.005016 | 0.850540 |
| HCl-SF$_6$ | 4.059816 | 0.078801 | 1.875646 | 0.032268 | 0.921199 |

*2.2 Spectra analysis*

To analyse the measured spectra, we used the Hartmann-Tran (HT) profile [1,2], the β-corrected HT (βHT) [28,29] and the speed dependent (complex) hard-collision model (SDcHC) [31-33] as well as some more simplified ones. For a given relative wavenumber $\Delta\sigma$, the HT is function of seven line-shape parameters, $\phi_{HT} = f(\Gamma_D, \Gamma_0, \Gamma_2, \Delta_0, \Delta_2, \nu_{VC}, \eta)$ where $\Gamma_D$ is the Doppler line width, $\Gamma_0$, $\Delta_0$ and $\Gamma_2$, $\Delta_2$ are respectively the speed-averaged collision-induced line broadening and shifting and their speed dependence components, $\nu_{VC}$ is the velocity-changing frequency and $\eta$ is the correlation parameter. A detailed description of the profile can be found in Ref. [1]. The functional form of the βHT profile can be obtained by replacing $\nu_{VC}$ in the HT by $\beta_\alpha(\chi)\nu_{VC}$ [28,29], i.e.:

$$\phi_{\beta HT} = f(\Gamma_D, \Gamma_0, \Gamma_2, \Delta_0, \Delta_2, \beta_\alpha(\chi)\nu_{VC}, \eta), \quad (1)$$

with $\beta_\alpha(\chi)$ the *β*-correction. The latter was determined in order to make the HT profile as close as possible to the speed-dependent billiard-ball (SDBB) profile [40] in which the velocity-changing collisions are described by the billiard-ball (BB) model [40-42] since it was shown that the BB model led to a better description of collision-induced velocity changes than the HC model for H$_2$ [27]. This correction depends on the perturber-to-absorber mass ratio $\alpha = m_p/m_a$ and the $\chi = \nu_{VC}/\Gamma_D$ ratio by the following expression [29]:

$$\beta_\alpha(\chi) = A_\alpha \tanh(B_\alpha \log_{10}\chi + C_\alpha) + D_\alpha, \quad (2)$$

where $A_\alpha, B_\alpha, C_\alpha$ and $D_\alpha$ are analytical α-dependent parameters given respectively by Eqs. (7a-7d) of Ref. [29]. The values of these parameters for the considered molecular systems are listed in **Table 2**.

The SDcHC model has the same number of parameters as the HT profile. Recall that in addition of 4 parameters describing the speed dependent line width and shift (i.e. $\Gamma_0$, $\Delta_0$ and $\Gamma_2$, $\Delta_2$) the HT profile assumes a partial correlation between velocity changing and internal state changing collisions (through the correlation parameter η) while the Dicke narrowing effect is characterized by a real-valued parameter ($\nu_{VC}$). In the SDcHC model, there is no correlation parameter but the Dicke narrowing effect is described by a complex number ($\nu_{VC} = \nu_{VC}^r + i\nu_{VC}^i$) with two collisional parameters $\nu_{VC}^r$ and $\nu_{VC}^i$. The functional form of the SDcHC model can thus be easily derived by setting the correlation parameter η to zero and replacing the velocity-changing frequency $\nu_{VC}$ by its complex form $\nu_{VC}^r + i\nu_{VC}^i$ in that of the HT profile, i.e.:

$$\phi_{SDcHC} = f(\Gamma_D, \Gamma_0, \Gamma_2, \Delta_0, \Delta_2, \nu_{VC}^r + i\nu_{VC}^i, 0). \quad (3)$$

Note that when there is no correlation in the HT profile and when the complex component of the Dicke narrowing parameter is set to zero in the SDcHC model, these two models are identical and become the SDHC model.

In addition to the HT, βHT and the SDcHC we also used some limiting cases of these profiles in the spectra analysis, i.e. the speed dependent hard collision profile, SDHC, the SDHC profile together with the β-correction (the βSDHC) (by setting $\eta = 0$ in the HT and βHT, respectively); the hard collision profile, HC (setting $\eta = \Gamma_2 = \Delta_2 = 0$ in the HT), the speed dependent Voigt, SDV (with $\eta = \nu_{VC} = 0$ in the HT) and the Voigt profile, V (setting $\eta = \Gamma_2 = \Delta_2 = \nu_{VC} = 0$ in the HT).

We used a multi-spectrum fitting procedure to analyse the measured spectra, i.e. measurements at various pressures were simultaneously adjusted. For each studied transition, the broadening coefficient $\gamma_0$ ($\Gamma_0/P$ with P the pressure) and its speed dependent component $\gamma_2$ ($\Gamma_2/P$), the speed dependent component $\delta_2$ ($\Delta_2/P$) of the line shifting, the real $\beta_r$ ($\nu_{VC}^r/P$) and the imaginary $\beta_i$ ($\nu_{VC}^i/P$) parts of the velocity-changing collision frequency, and the correlation parameter η were constrained to be the same for all pressures. The position of the line maximum (including the non-perturbed position and the pressure shift $\Delta_0$) was fit in each spectrum. Because of possible adsorption/desorption of HCl in the cell and cylinder walls, the partial pressure of HCl in each sample cannot be determined with precision, and thus the line integrated area was fit for each pressure. In addition, a linear base line was also adjusted for each considered spectrum.

## 3. Results and discussions

**Figure 1** presents the apparent half-width at half-maxima (HWHM) of the considered spectra of HCl in He, $SF_6$ and Ar versus pressure. As already pointed out in [39] for HCl-Ar and observed in this figure for HCl-He and HCl-$SF_6$, the Dicke narrowing effect is strong for all considered situations and more pronounced for the R(9) line. The influence of this effect on the HCl spectral shapes increases from He to $SF_6$ and it is more conspicuous for Ar. The same situation was reported for HCl diluted in Ne, Xe and Ar [35] which was explained by the fact that an equal-mass perturber is most effective in changing the velocity through collisions [35]. It is clear that a precise modeling of the Dicke narrowing effect is needed to correctly represent the measured spectra of HCl, especially for HCl diluted in Ar. The results obtained with the use of the HT, βHT and SDcHC profiles to fit the measured spectra are presented in the next sections.

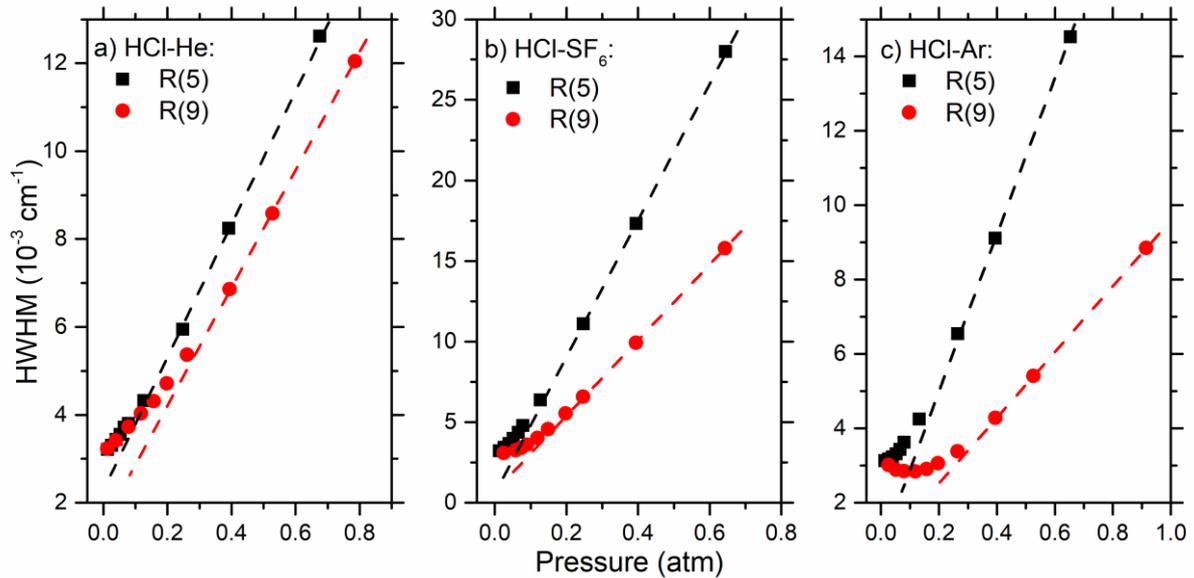

**Figure 1**. The apparent HWHM of the measured absorption spectra of the (1-0) R(5) and R(9) lines of HCl broadened by He, $SF_6$ and Ar vs pressure. The dash-straight lines show the linear fits of the values of the HWHM at the two highest pressures.

*3.1 Fit residuals*

For each collision-partner and each line, the HCl spectra measured at various pressures were first normalized by their maximum absorptions and then simultaneously adjusted using the above-mentioned profiles (**Sec. 2.2**). A multi-spectrum fitting procedure reduces the numerical correlations between line-shape parameters. Spectral ranges of about thirty times the corresponding apparent HWHM were used in the fits in order to have the same ratio of the HWHM to the total spectral range for each considered pressure. Note that in [39], the measured spectra of the R(5) and R(9) lines of HCl-Ar were analysed but with a single spectrum fitting procedure and only the obtained values of the pressure shift and width were reported, there is therefore no repetition in the following results of this work.

**Figures 2-4** present the residuals obtained from fits of the measured spectra with the different line-shape profiles. Comparison between the obtained fit residuals can be obtained by considering the quality of the fit (QF), defined as the ratio of the maximum absorption to the root mean square of the fit [9]. The QF is thus equivalent to the signal-to-noise ratio in the case

of a fit with perfectly adapted line shape. As can be observed in **Figs. 2-4**, for each considered collisional-partners and each line-shape model, the QF obtained for the R(5) line is always higher than that for the R(9), especially for HCl-Ar. This is due to a higher signal-to-noise ratio of the measured spectra of the R(5) line as well as a better representation of this line by the models. For all considered transitions and collision-partners, the V leads to large deviation with respect to the measured profile, with fit residuals as large as 6%, 5%, 2% and 14%, 10%, 3% for R(5) and R(9) transitions of HCl diluted in Ar, $SF_6$ and He, respectively. Non-Voigt effects are thus stronger for R(9) than R(5) and for HCl-Ar and HCl-$SF_6$ than for HCl-He. This result is consistent with **Fig. 1**.

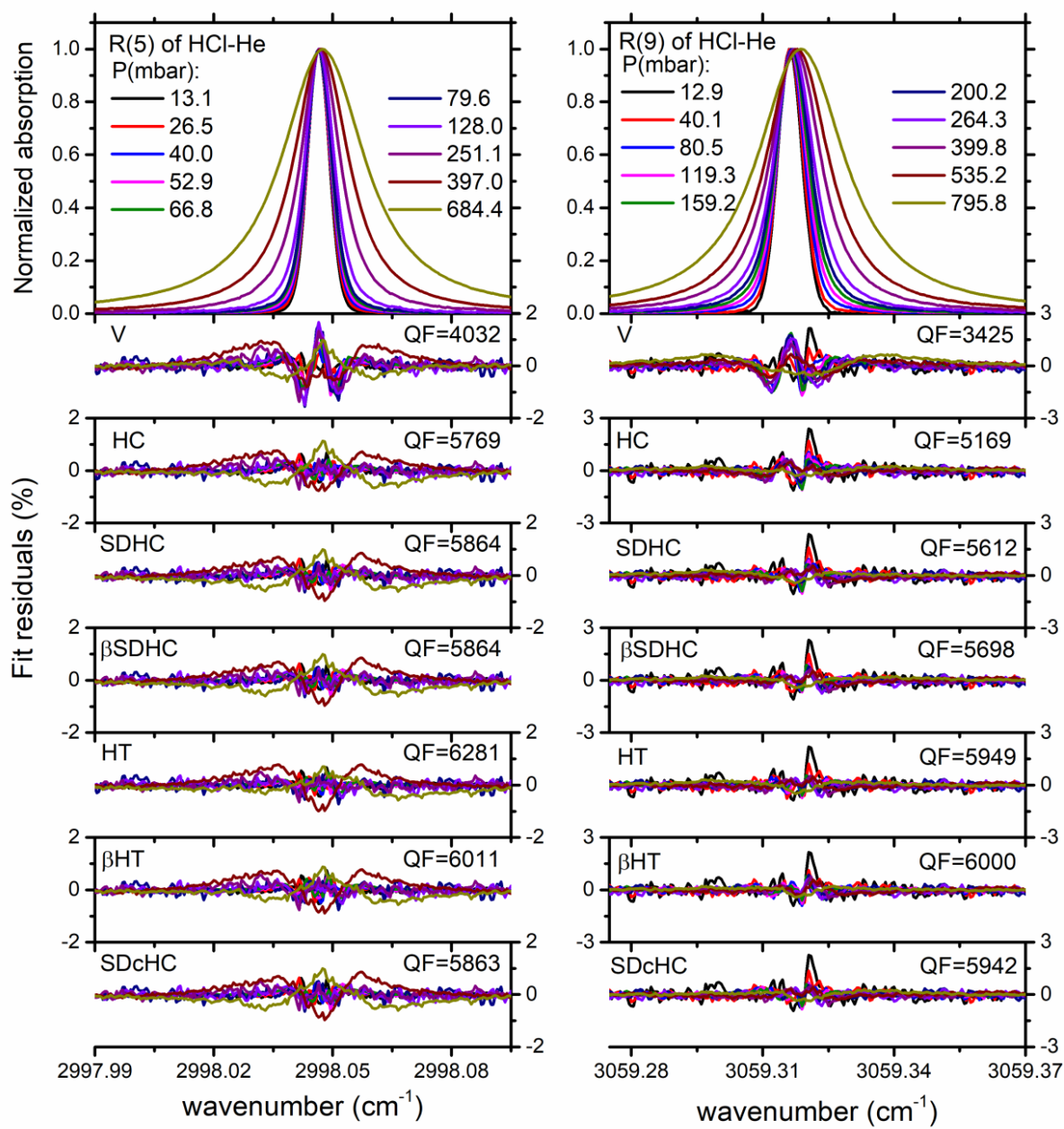

**Figure 2**: Measured absorption spectra of the HCl (1-0) R(5) (left) and R(9) (right) transitions broadened by He, normalized by the maximum absorption and the corresponding residuals obtained from fits of these spectra with various profiles.

Note that for all considered lines and perturbers, considering only the Dicke narrowing effect (by using the HC model) leads to significantly better fits of measured spectra than the SDV profile (residuals not shown in **Figs. 2-4** for clarity). This result is due to the strong velocity changing effect affecting the shape of the considered lines. Note that this situation is in opposite with what usually observed for many other molecular systems (e.g. self-broadened $C_2H_2$ [43], self-broadened $CO_2$ [13], air-broadened $CO_2$ [9], $N_2$-broadened $O_2$ [16], $H_2O$ perturbed by $N_2$ and $SF_6$ [12] and $CH_4$ perturbed by $N_2$ [18]) for which the SDV profile leads to much better agreement with measurements than models taking only the Dicke narrowing effect into account. Asymmetric features are clearly observed for the R(9) line (see **Figs. 2-4**), especially at low pressures while they are much weaker for the R(5) line. This characteristic was also reported for several high rotational quantum number lines ($J \geq 6$) of HF and HCl diluted in $N_2$ and air [36] and for the P(14) line of HCl broadened by Xe [37].

As can be observed in **Fig. 2**, Dicke narrowing plays a rather important role in the observed non-Voigt effects for HCl-He, since using the HC model significantly improves the fit residuals and the QF with respect to results obtained with the Voigt profile. The results are slightly improved when adding the speed dependence effect using the SDHC model. For both lines, adding the β correction to the SDHC or the HT models does not change the results. This shows that the hard collision model is sufficient to model Dicke narrowing effect for He broadened HCl lines. For the R(5) line, using the HT model leads to the best fit residuals and QF while for the R(9) line, the HT and SDcHC give similar fit residuals and QFs. For the R(5) line, using the SDcHC model with the complex Dicke narrowing parameter leads to similar results than those obtained with the SDHC, adding the imaginary part to the Dicke narrowing parameter therefore has no effect on the spectra.

Larger non-Voigt effects are observed for HCl-$SF_6$ (**Fig. 3**). The use of the SDHC model significantly reduces the fit residuals with respect to the Voigt profile. Adding the correlation parameter by the use of the HT profile improves significantly the quality of the fit with respect to the use of the SDHC. The β correction does not change the fit residuals for the R(5) line, it deteriorates the quality of the fit with the HT profile. However, this β correction does have a great influence for the R(9) line for which the Dicke narrowing effect is strong (see **Fig. 1**). It shows that the β correction is indeed relevant for systems heavily affected by the Dicke narrowing effect. For the two considered lines, the SDcHC model does a better job than the HT profile.

For HCl-Ar (**Fig. 4**), using the β correction with the SDHC profile leads to the best fits of the measured spectra for both the R(5) and R(9) lines. This is consistent with the large Dicke narrowing effect observed in Fig. 1 for HCl-Ar. The SDHC and the HT give similar results showing that adding the correlation parameter is not necessary for this system. Compared with the SDHC profile, its complex version SDcHC leads to similar fit residual and QF for the R(5) line but to better results for the R(9) line.

In order to go further, we used the β correction along with the SDcHC model to fit all considered spectra (the β correction is applied to the real component of the complex Dicke narrowing). The obtained results demonstrate that this βSDcHC model leads to the same quality of fits, fit residuals and retrieved line parameters as the SDcHC model for the cases of the R(5) line broadened by He, $SF_6$ and Ar and of the R(9) line of HCl in He. For the R(9) line of HCl in $SF_6$ and in Ar for which the Dicke narrowing is the most important, as expected, using the βSDcHC model leads to better fit residuals and QFs (see Fig. 5).

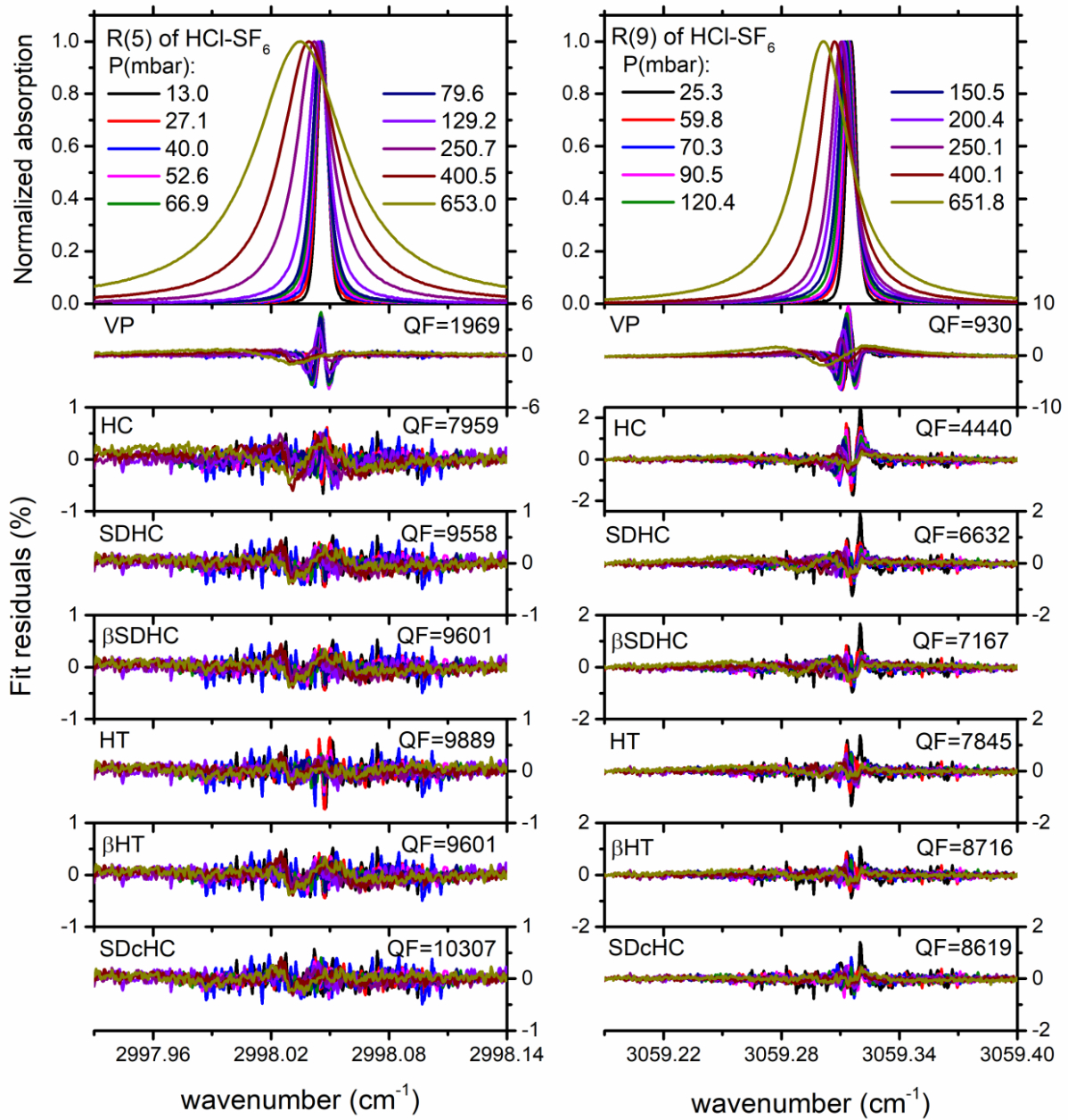

**Figure 3**: The same as **Fig. 2** but for the HCl (1-0) R(5) and R(9) transitions perturbed by $SF_6$

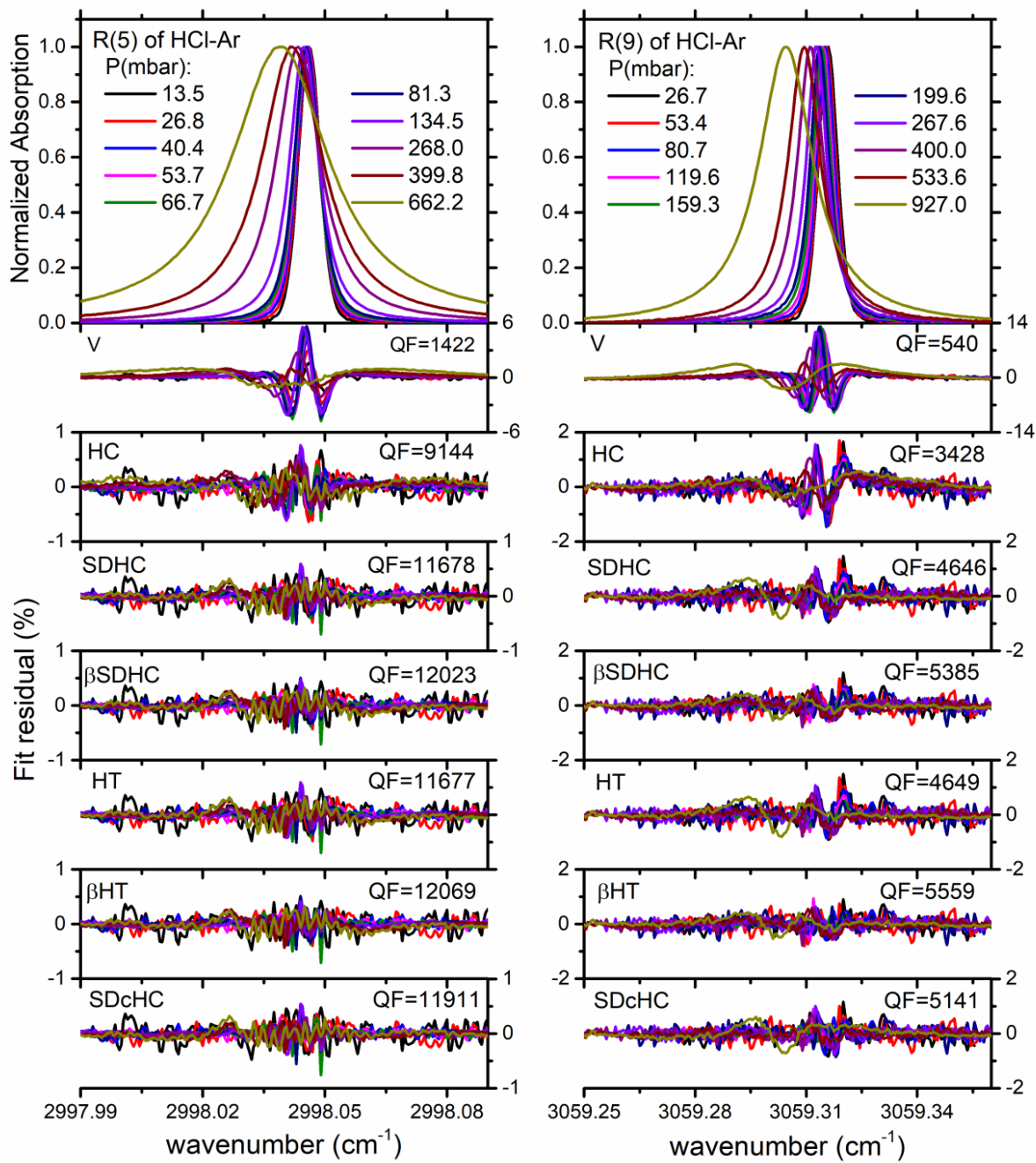

**Figure 4**: The same as **Fig. 2** but for the HCl (1-0) R(5) and R(9) transitions broadened by Ar

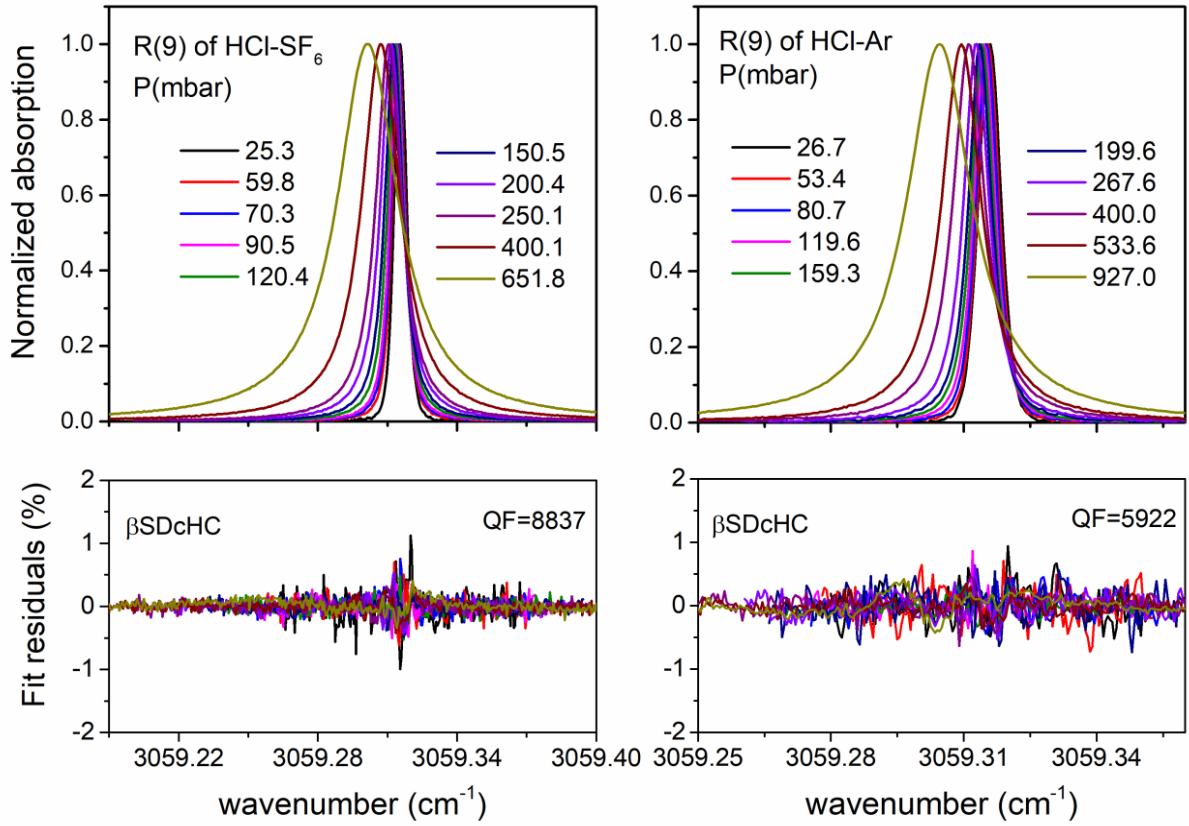

**Figure 5**: Measured absorption spectra of the R(9) line of HCl broadened by $SF_6$ (left) and Ar (right) and the corresponding residuals obtained from fits of these spectra with the βSDcHC model.

*3.2 Fit parameters*

The various line-shape parameters and their uncertainties obtained from fits of the measured spectra with all the considered models are given in **Tables 3-5**, for HCl-He, HCl-$SF_6$ and HCl-Ar, respectively. Since the partial pressures of HCl are not known with high precision, the values of the absolute line intensity are not reported here. The quoted uncertainties in **Tables 3-5** correspond to the combined uncertainties which include uncertainties due to those of pressures, temperatures and statistical uncertainties of the corresponding parameters obtained from the fits.

**Table 3:** Line-shape parameters for the HCl (1-0) R(5) and R(9) lines broadened by He obtained from fits of the measured spectra with different line-shape models. All parameters ($\delta_0$, $\gamma_0$, $\gamma_2$, $\delta_2$ and $\beta_r$, $\beta_i$) are given in $10^{-3}$ cm$^{-1}$atm$^{-1}$. Parameters obtained with the Galatry profile (soft collision model, SC [44]) are also reported. The quoted uncertainties (between parentheses) are in the same units of the last digit of the corresponding values.

| Line | Models | $\delta_0$ | $\delta_0^{Ref}$ | $\gamma_0$ | $\gamma_0^{Ref}$ | $\gamma_2$ | $\delta_2$ | $\beta_r$ | $\beta^{Ref}$ | $\beta_i$ | $\eta$ |
|---|---|---|---|---|---|---|---|---|---|---|---|
| R(5) | V | 1.63(7) | 2.0[45] | 17.77(3) | 22.5[45] | | | | | | |
| | SDV | 1.64(7) | | 18.23(2) | | 1.41(1) | -0.22(1) | | | | |
| | HC | 1.63(7) | | 18.03(3) | | | | 2.90(4) | 6.1(1)[37], 5.4[37],* | | |
| | SC | 1.63(7) | | 18.08(3) | | | | 3.82(7) | 7.6(1)[37] | | |
| | SDHC | 1.64(7) | | 18.23(3) | | 1.40(1) | -0.22(1) | 0.03(5) | | | |
| | βSDHC | 1.63(7) | | 18.23(3) | | 1.40(2) | -0.23(3) | 0.05(9) | | | |
| | HT | 1.63(7) | | 18.38(2) | | -4.20(4) | -0.23(3) | 20.77(5) | | | 0.634(2) |
| | βHT | 1.64(7) | | 18.13(2) | | -1.43(2) | -0.41(3) | 6.24(9) | | | -0.123(3) |
| | SDcHC | 1.64(7) | | 18.23(3) | | 1.40(1) | -0.26(1) | 0.02(5) | | 0.10(4) | |
| R(9) | V | 3.21(4) | 3.3[45] | 13.95(1) | 17.0[45] | | | | | | |
| | SDV | 3.35(5) | | 14.43(2) | | 1.29(1) | 0.36(1) | | | | |
| | HC | 3.21(4) | | 14.25(2) | | | | 2.86(4) | 6.1(1)[37], 5.4[37],* | | |
| | SC | 3.21(4) | | 14.31(2) | | | | 3.89(4) | 7.6(1)[37] | | |
| | SDHC | 3.35(5) | | 14.35(2) | | 0.91(1) | 0.39(1) | 0.90(4) | | | |
| | βSDHC | 3.20(5) | | 14.29(1) | | -0.73(2) | 0.61(2) | 6.99(5) | | | |
| | HT | 3.30(5) | | 14.38(1) | | -0.37(2) | 1.67(3) | 25.76(4) | | | 1.580(3) |
| | βHT | 3.20(5) | | 14.35(2) | | -0.33(3) | 0.78(3) | 13.41(7) | | | 0.425(4) |
| | SDcHC | 3.20(5) | | 14.37(1) | | 1.11(2) | -0.47(2) | 0.40(4) | | 2.42(3) | |

Note: "*" means a calculated value.

**Table 4:** Line-shape parameters for the HCl (1-0) R(5) and R(9) lines of HCl broadened by $SF_6$, obtained from fits of the measured spectra with different line-shape models. All parameters ($\delta_0$, $\gamma_0$, $\gamma_2$, $\delta_2$ and $\beta_r$, $\beta_i$) are given in $10^{-3}$ cm$^{-1}$atm$^{-1}$. The quoted uncertainties (between parentheses) are in the same units of the last digit of the corresponding values.

| Line | Model | $\delta_0$ | $\gamma_0$ | $\gamma_2$ | $\delta_2$ | $\beta_r$ | $\beta^{Ref}$ | $\beta_i$ | $\eta$ |
|---|---|---|---|---|---|---|---|---|---|
| $SF_6$ | V | -18.16(6) | 41.94(5) | | | | | | |
| | SDV | -18.66(5) | 44.86(5) | 8.65(2) | -1.21(3) | | | | |
| | HC | -18.16(6) | 42.87(5) | | | 25.35(10) | 54.6* | | |
| | SC | -18.16(6) | 42.99(5) | | | 32.22(11) | | | |
| | SDHC | -18.52(5) | 43.39(5) | 4.50(2) | -1.55(2) | 15.40(9) | | | |
| | βSDHC | -18.46(6) | 43.39(5) | 4.47(2) | -1.55(2) | 17.94(9) | | | |
| | HT | -18.88(5) | 43.07(5) | -3.32(2) | -4.27(2) | 117.00(15) | | | 2.142(2) |
| | βHT | -18.46(6) | 43.39(5) | 4.50(4) | -1.57(3) | 18.90(69) | | | 0.020(2) |
| | SDcHC | -18.66(5) | 43.27(5) | 4.63(2) | -2.79(2) | 15.13(8) | | 4.96(7) | |
| $SF_6$ | V | -22.77(6) | 22.34(4) | | | | | | |
| | SDV | -22.23(7) | 27.00(4) | 8.86(3) | 1.04(3) | | | | |
| | HC | -22.77(6) | 24.12(3) | | | 31.44(13) | | | |
| | SC | -22.77(6) | 24.29(3) | | | 38.71(15) | | | |
| | SDHC | -22.72(6) | 23.75(3) | 0.34(2) | 3.71(2) | 29.64(9) | 54.6* | | |
| | βSDHC | -22.85(5) | 23.73(3) | -0.35(2) | 4.01(2) | 34.26(9) | | | |
| | HT | -22.67(5) | 24.26(3) | 13.32(4) | 5.33(4) | 36.49(9) | | | 0.978(1) |
| | βHT | -22.67(5) | 24.27(3) | 11.21(5) | 6.00(4) | 43.09(11) | | | 0.971(1) |
| | SDcHC | -22.85(5) | 24.82(3) | 4.42(1) | 0.52(1) | 20.62(7) | | 9.29(6) | |
| | βSDcHC | -22.85(5) | 24.47(3) | -3.38(2) | -0.41(2) | 41.89(4) | | 12.18(7) | |

Note: "*" means a calculated value

**Table 5:** Line-shape parameters for the HCl (1-0) R(5) and R(9) lines of HCl broadened by Ar, obtained from fits of the measured spectra with different line-shape models. All parameters ($\delta_0$, $\gamma_0$, $\gamma_2$, $\delta_2$ and $\beta_r$, $\beta_i$) are given in $10^{-3}$ cm$^{-1}$atm$^{-1}$. The quoted uncertainties (between parentheses) are in the same units of the last digit of the corresponding values.

| Line | Model | $\delta_0$ | $\delta_0^{Ref}$ | $\gamma_0$ | $\gamma_0^{Ref}$ | $\gamma_2$ | $\delta_2$ | $\beta_r$ | $\beta^{Ref}$ | $\beta_i$ | $\eta$ |
|---|---|---|---|---|---|---|---|---|---|---|---|
| R(5) | V | -11.15(9) | -12.9[45], -12.1[46], -11.1[47], -13.0[48], -11.15[39] | 20.70(3) | 24.5[45], 20.7[46], 21.8[46], 24.2[48], 21.03[39] | | | | | | |
|  | SDV | -11.09(10) | | 23.55(3) | | 6.14(2) | 0.07(2) | | | | |
|  | HC | -11.15(9) | | 21.72(3) | | | | 16.12(5) | 13.7(5) [34], 23.4(20) [49], 23.2[37],* | | |
|  | SC | -11.15(9) | | 21.88(3) | | | | 21.57(7) | 16.2(3) [37], 24.5(2) [37] | | |
|  | SDHC | -11.09(10) | | 22.27(3) | 21.74[39] | 3.28(1) | 0.17(2) | 9.44(4) | | | |
|  | βSDHC | -11.09(10) | | 22.22(3) | | 3.01(1) | 0.21(1) | 13.49(5) | | | |
|  | HT | -11.09(10) | | 22.27(3) | | 3.25(1) | 0.17(1) | 9.10(12) | | | -0.016(7) |
|  | βHT | -11.09(10) | | 22.21(2) | | 3.15(1) | 0.23(1) | 16.75(21) | | | 0.118(9) |
|  | SDcHC | -11.27(10) | | 22.28(3) | | 3.32(1) | -0.27(1) | 9.33(4) | | 1.62(3) | |
| R(9) | V | -12.78(3) | -15.2[45], -15.0[48], -13.07[39] | 7.64(3) | 11.0[45], 11.1[48], 8.8[35], 8.47[39] | | | | | | |
|  | SDV | -12.57(4) | | 11.66(2) | | 5.48(2) | 0.31(2) | | | | |
|  | HC | -12.78(3) | | 9.24(1) | | | | 19.49(8) | 13.7(5) [34], 23.4(20) [49], 23.2[37],* | | |
|  | SC | -12.78(3) | | 9.40(1) | | | | 24.91(8) | 16.2(3) [37], 24.5(2) [37] | | |
|  | SDHC | -12.49(4) | | 10.03(1) | 9.68[39] | 3.51(1) | 0.88(1) | 12.58(5) | | | |
|  | βSDHC | -12.51(4) | | 9.92(1) | | 3.23(1) | 1.01(1) | 16.36(6) | | | |
|  | HT | -12.49(4) | | 10.04(1) | | 3.33(1) | 0.84(1) | 11.40(6) | | | -0.123(4) |
|  | βHT | -12.49(4) | | 9.83(1) | | 4.20(1) | 1.35(1) | 21.59(5) | | | 0.517(2) |
|  | SDcHC | -12.77(5) | | 10.14(2) | | 3.64(1) | 0.35(1) | 12.17(5) | | 2.91(5) | |
|  | βSDcHC | -12.77(5) | | 10.09(1) | | 3.55(1) | 0.40(1) | 15.52(6) | | 2.70(4) | |

Note: "*" means a calculated value

Comparisons between the values of various line-shape parameter ($\delta_0$, $\gamma_0$, $\gamma_2$, $\delta_2$ and $\beta_r$) obtained from fits with different models shown in **Figure 6**. For clarity, these parameters have been divided by the corresponding values obtained with the SDcHC model. As can be observed in **Fig. 6a,** as well as in **Tables 3-5**, the pressure shift coefficients $\delta_0$ are rather independent of the used line-shape models. The values of the broadening coefficients (**Fig. 6b** and **Tables 3-5**) obtained with all line-shape models which take into account both the Dicke narrowing and the speed dependence effects (i.e. the βHT, HT, βSDHC, SDHC and SDcHC) are almost identical and they are larger than those obtained with the HC. The SDV leads to the largest values of $\gamma_0$ while those obtained with the V are significantly smaller (up to about 20%). The obtained values $\delta_0$ and $\gamma_0$ of this work are in rather good agreement with data from other studies for HCl lines in Ar (see **Table 4**) but for HCl lines in He (**Table 3**) they differ up to 26.6% from those measured by Rank et al [45]. To the best of our knowledge, there is no available data in the literature for HCl perturbed by $SF_6$.

For the other line-shape parameters (of higher order), much larger differences are observed for different line-shape models used. The largest difference between the SDcHC model and the others is observed for the speed-dependent components of the line-shift $\delta_2$. This parameter is quite small for the R(5) line but it becomes significantly larger for the R(9) transition, corresponding to the larger asymmetric features of the R(9) line. As can be seen in **Tables 3-5**, the Dicke narrowing coefficients obtained from fits with the HC ($\beta^{HC}$) and SC ($\beta^{SC}$) depend not only on the buffer gas but also on the considered line. For all perturbers, $\beta^{SC}$ is always larger than $\beta^{HC}$ with the ratios $\beta^{SC}/\beta^{HC}$ varying from 1.23 to 1.36. This observation is consistent with the measured values of Hurtmans et al. [37] for HCl in Xe, He and $N_2$ and also with the theoretical prediction of Wojtewicz et al. [50] in which it was shown that, from low to high pressure limits, $\beta^{SC}/\beta^{HC}$ ratios vary from $\frac{3}{2}(\pi - 2)$ to 1. For HCl diluted in Ar, the obtained values of $\beta^{HC}$ and $\beta^{SC}$ are in good agreements with those of other studies realized for other lines as well as with the calculated value of the HCl-Ar diffusion coefficient $\beta^{Diff}$ (**Table 5**). For HCl in He, our $\beta^{HC}$ and $\beta^{SC}$ values of the considered lines are nearly two times smaller than the corresponding parameter determined from the diffusion coefficient and are 2.6 times smaller than measured values of the P(14) line in [37]. **Figs. 6d-e** respectively show the comparisons for $\gamma_2$ and $\beta_r$. The difference between line parameters obtained from simplified models (i.e. SDV, HC and SC) and those retrieved with more sophisticated ones (i.e. SDHC, βSDHC, HT, βHT and SDcHC) is due to the fact that the first ones are only considering either the speed-dependence effect or the Dicke narrowing one. However, it is worth noting that for many situations, the obtained values for these refined line-shape parameters are only effective, because of the strong numerical correlation between them. For instance, with both the SDHC and βSDHC, the correlation factor between β and $\gamma_2$ for the R(5) transition of HCl in He is 0.88, very close to 100% correlation.

Note that using the βSDcHC model leads to exactly the same retrieved parameters as with the SDcHC profile except for the R(9) line in $SF_6$ and Ar for which the corresponding line parameters are reported in Tables 4 and 5. While these parameters seem to be properly determined for the case of the R(9) line of HCl in Ar, using the βSDcHC to fit the measured spectra of the R(9) line of HCl in $SF_6$ leads to a negative hence unphysical value of $\gamma_2$ (see Table 4).

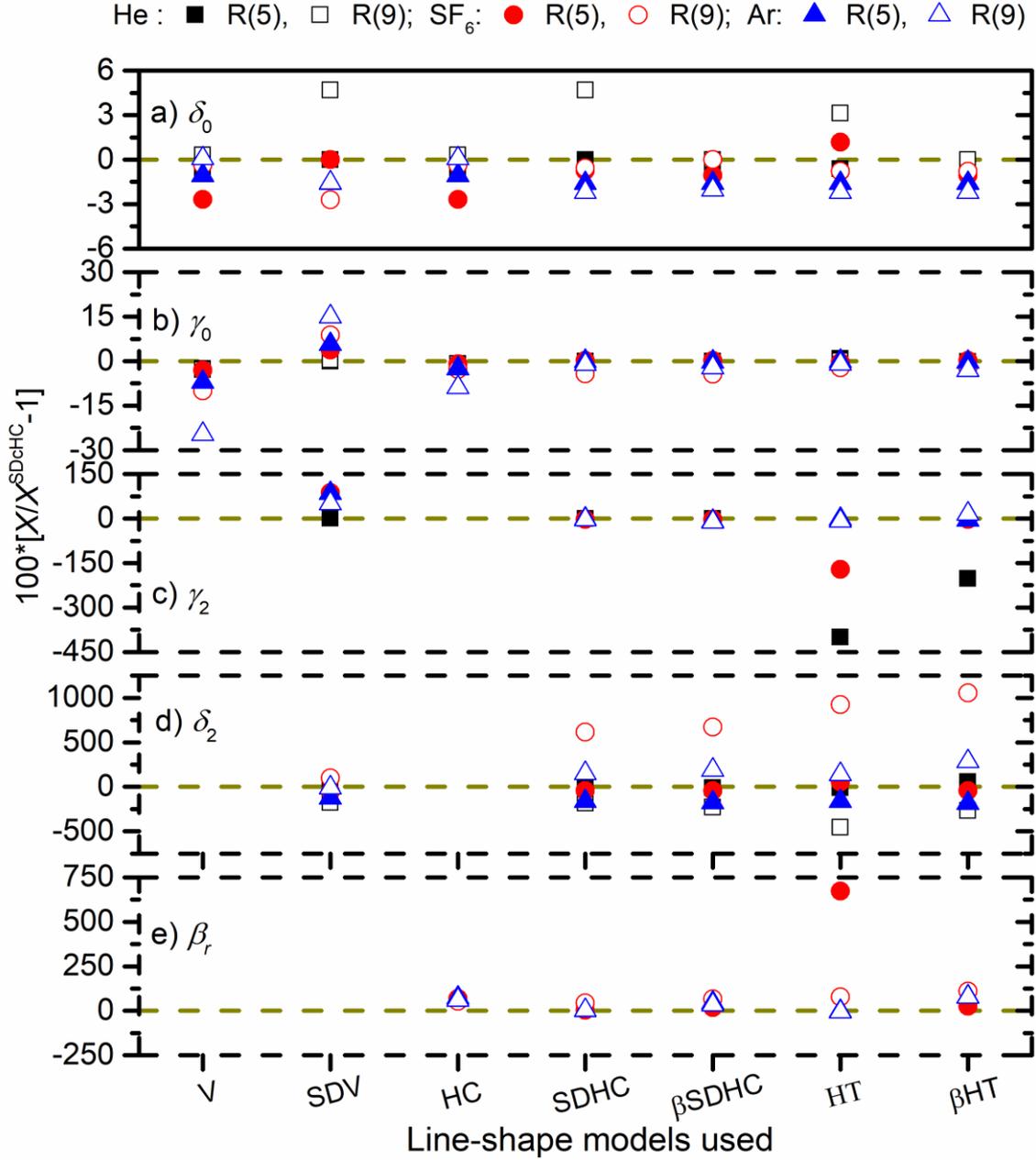

**Figure 6**. Relative differences between the obtained values of the line parameters (i.e., $\delta_0$, $\gamma_0$, $\gamma_2$, $\delta_2$ or $\beta_r$) from fits of the measured spectra with the different line-shape models (V, SDV, HC, SDHC, βSDHC, HT and βHT) and those from fit with the SDcHC for R(5) and R(9) transitions of HCl perturbed by He, SF$_6$ and Ar (black solid- and open- squares, red solid- and open- circles and blue solid- and open- triangles, respectively).

As shown in the previous section, the HT and the SDcHC profiles lead to similar QFs which are better than the simplified models as HC and SDHC, except in situations where the Dicke narrowing effect is significant for which the use of the β-correction can greatly improve the quality of the fit. While the values of the broadening and shifting coefficients obtained with these two profiles are very close to each other, the refined line-shape parameters show large differences. In particular, with the HT profile, for some sets of the measured spectra, the fitted value of η can be negative or larger than 1 and the values of $\gamma_2$ are sometimes negative, which have no physical meaning. That is due to the strong correlation between η, $\gamma_2$ and $\beta_r$ in the HT profile [1]. In order to go further, in **Table 6** we show the correlation factors $[f_{cor}(X,Y)]$ between the various line-shape parameters obtained from fits with the HT and the SDcHC models, for all considered lines and perturbers. As can be observed in this table, with the HT profile, the values of $f_{cor}(\eta,\beta_r)$ for the R(5) and R(9) lines of HCl-He are respectively 0.95 and 0.99 while it is 0.98 for the R(5) line of HCl in $SF_6$ and HCl in Ar. These situations indeed correspond to those where the value of $\eta$ is negative or larger than 1. In addition, the correlations between $\eta$ and $\gamma_2$ and between $\beta_r$ and $\gamma_2$ are also important. With the SDcHC model, the obtained $f_{cor}(X,Y)$ factors are much smaller, namely for the case of HCl-$SF_6$ and HCl-Ar. For HCl-He, the value of $f_{cor}(\beta_r,\gamma_2)$ obtained with the SDcHC model is larger than that obtained with the HT profile which is consistent with the fact that the HT profile gives better quality of fit then the SDcHC model in this case. However, in overall, the use of the SDcHC model leads to better correlation factors and thus enables to more properly determine the line-shape parameters.

|  |  | HT |  |  |  |  | SDcHC |  |  |  |  |
|---|---|---|---|---|---|---|---|---|---|---|---|
|  | Pars. | $\gamma_0$ | $\gamma_2$ | $\delta_2$ | $\beta_r$ | $\eta$ | Pars. | $\gamma_0$ | $\gamma_2$ | $\delta_2$ | $\beta_r$ | $\beta_i$ |
| HCl-He | $\gamma_0$ | 1.00 | -0.32 | -0.75 | -0.62 | 0.55 | $\gamma_0$ | 1.00 | -0.74 | -0.18 | -0.51 | -0.02 |
|  | $\gamma_2$ | 0.25 | 1.00 | 0.00 | 0.50 | -0.45 | $\gamma_2$ | -0.74 | 1.00 | 0.00 | 0.90 | -0.09 |
|  | $\delta_2$ | -0.04 | 0.00 | 1.00 | 0.44 | -0.39 | $\delta_2$ | -0.05 | 0.00 | 1.00 | 0.08 | 0.90 |
|  | $\beta_r$ | -0.49 | 0.66 | 0.02 | 1.00 | -0.99 | $\beta_r$ | -0.44 | 0.88 | 0.04 | 1.00 | 0.00 |
|  | $\eta$ | 0.70 | -0.36 | -0.01 | -0.94 | 1.00 | $\beta_i$ | 0.02 | -0.04 | 0.88 | 0.00 | 1.00 |
|  | Pars. | $\gamma_0$ | $\gamma_2$ | $\delta_2$ | $\beta_r$ | $\eta$ | Pars. | $\gamma_0$ | $\gamma_2$ | $\delta_2$ | $\beta_r$ | $\beta_i$ |
| HCl-$SF_6$ | $\gamma_0$ | 1.00 | -0.66 | 0.05 | -0.50 | 0.76 | $\gamma_0$ | 1.00 | -0.89 | -0.08 | -0.38 | -0.13 |
|  | $\gamma_2$ | -0.58 | 1.00 | 0.00 | 0.91 | -0.71 | $\gamma_2$ | -0.78 | 1.00 | 0.00 | 0.69 | 0.10 |
|  | $\delta_2$ | 0.70 | 0.00 | 1.00 | -0.24 | -0.31 | $\delta_2$ | -0.32 | 0.00 | 1.00 | -0.10 | 0.69 |
|  | $\beta_r$ | -0.44 | 0.26 | -0.22 | 1.00 | -0.45 | $\beta_r$ | -0.31 | 0.62 | 0.19 | 1.00 | 0.00 |
|  | $\eta$ | 0.40 | -0.35 | 0.07 | -0.98 | 1.00 | $\beta_i$ | 0.10 | -0.19 | 0.62 | 0.00 | 1.00 |
|  | Pars. | $\gamma_0$ | $\gamma_2$ | $\delta_2$ | $\beta_r$ | $\eta$ | Pars. | $\gamma_0$ | $\gamma_2$ | $\delta_2$ | $\beta_r$ | $\beta_i$ |
| HCl-Ar | $\gamma_0$ | 1.00 | -0.93 | 0.12 | -0.31 | 0.70 | $\gamma_0$ | 1.00 | -0.94 | -0.04 | -0.28 | -0.10 |
|  | $\gamma_2$ | -0.84 | 1.00 | 0.00 | 0.50 | -0.87 | $\gamma_2$ | -0.84 | 1.00 | 0.00 | 0.50 | 0.09 |
|  | $\delta_2$ | 0.02 | 0.00 | 1.00 | -0.20 | -0.01 | $\delta_2$ | -0.06 | 0.00 | 1.00 | -0.09 | 0.50 |
|  | $\beta_r$ | -0.35 | 0.72 | -0.02 | 1.00 | -0.85 | $\beta_r$ | -0.35 | 0.71 | 0.02 | 1.00 | 0.00 |
|  | $\eta$ | 0.49 | -0.83 | 0.01 | -0.98 | 1.00 | $\beta_i$ | 0.01 | -0.02 | 0.71 | 0.00 | 1.00 |

**Table 6.** Numerical correlation factors between various line-shape parameters obtained from fits with the HT and SDcHC profiles of the measured spectra of the R(5) (grey) and R(9) (yellow) lines of HCl diluted in He, $SF_6$ and Ar.

## 4. Conclusion

Spectra of the HCl (1-0) R(5) and R(9) lines broadened by He and $SF_6$ as well as Ar measured with a difference frequency laser spectrometer at room temperature and for large pressure ranges were used to test different refined line-shape models, including the Hartmann-Tran profile, the speed dependent complex hard collision model and the β-correction proposed for systems heavily affected by the Dicke narrowing effect. We showed that for all considered lines and perturbers, both the confinement narrowing and the speed dependences of the line broadening and shifting must be considered in order to correctly simulate the measured spectra. Using the β-correction together with the HT profile significantly improves the fit residuals except for HCl-He and for the R(5) line of HCl-SF$_6$, for which the influence of velocity changing collisions is smaller. The SDcHC model in which the Dicke narrowing coefficient is a complex number and thus characterized by two parameters (i.e. its real and imaginary components) leads to better quality of fit compared to the Hartmann-Tran profile, except for HCl-He. The results also show that numerical correlations between refined line-shape parameters of the HT profile are important and can lead to ill-determined parameters while they are more properly determined with the SDcHC model. Note that the latter can be easily obtained from the HT profile by adding an imaginary part to the Dicke narrowing coefficient and setting the correlation parameter to zero thus preserving all the advantages of the HT profile such as its low computer cost. This work thus confirms that the use of SDcHC for high resolution spectroscopy is more relevant than the HT profile.

**Acknowledgements**

The authors from Hanoi National University of Education is pleased to acknowledge the financial support of this research from the Vietnam National Foundation for Science and Technology Development (NAFOSTED) under grant number 103.03-2018.341. J.L.D. acknowledges the support from the MCIU project PID2020-113084GB-I00/AEI/10.13039/501100011033.